\documentclass[onecolumn]{article}%
\usepackage{amsmath}
\usepackage{amsfonts}
\usepackage{amssymb}
\usepackage{graphicx}%
\providecommand{\U}[1]{\protect\rule{.1in}{.1in}}
\begin{document}

\title{\hfill{\normalsize UMTG - 2}\bigskip\\Supersymmetric Biorthogonal Quantum Systems}
\author{Thomas Curtright$^{\text{\S ,\dag,}}${\small *}, Luca Mezincescu$^{\text{\S ,}%
}${\small *}, and David Schuster$^{\text{\S ,}}${\small *}\medskip\\$^{\text{\S }}${\small Department of Physics, University of Miami, Coral
Gables, Florida 33124}\\$^{\text{\dag}}${\small School of Natural Sciences, Institute for Advanced
Study, Princeton, New Jersey 08540}}
\maketitle

\begin{abstract}
We discuss supersymmetric biorthogonal systems, with emphasis given to the
periodic solutions that occur at spectral singularities of PT symmetric
models. \ For these periodic solutions, the dual functions are associated
polynomials that obey inhomogeneous equations. \ We construct in detail some
explicit examples for the supersymmetric pairs of potentials $V_{\pm}\left(
z\right)  =-U\left(  z\right)  ^{2}\pm z\frac{d}{dz}U\left(  z\right)  $ where
$U\left(  z\right)  \equiv\sum_{k>0}\upsilon_{k}z^{k}$.\ \ In particular, we
consider the cases generated by $U\left(  z\right)  =z$ and $z/\left(
1-z\right)  $. \ We also briefly consider the effects of magnetic vector
potentials on the partition functions of these systems.

\vfill

\noindent\underline
{\ \ \ \ \ \ \ \ \ \ \ \ \ \ \ \ \ \ \ \ \ \ \ \ \ \ \ \ \ \ \ \ \ \ \ \ \ \ \ \ \ \ \ \ \ \ \ \ \ \ \ \ \ \ \ }%

{\small *}curtright@physics.miami.edu, mezincescu@physics.miami.edu, \&
dschuster@physics.miami.edu \ 

\end{abstract}

\section{Introduction}

There has been some recent theoretical interest in non-hermitian
Schr\"{o}dinger equations, in the guise of \textquotedblleft PT symmetric
theories\textquotedblright\ \cite{BenderReview}. \ Several previous authors
have considered supersymmetry in this context \cite{SusyPT}. \ Here, we
consider a few elementary soluble examples and explore them in some detail.
\ We believe exact solvability permits the underlying structure to be
appreciated more completely. \ Some recent papers \cite{BiorthogonalPT,CM}
have also touched on the relevance of biorthogonal systems
\cite{BiorthogonalClassics} for PT symmetric models. \ As in \cite{CM}, we
wish again to stress the importance of such systems, and their generality,
only here in the context of supersymmetric examples. \ We are not aware of any
previous systematic discussion of supersymmetric biorthogonal systems along
the lines of that given here.

We consider models whose Hamiltonians are of the form%
\begin{gather}
H_{\pm}=\left(  z\frac{d}{dz}+\nu\mp U\right)  \left(  z\frac{d}{dz}+\nu\pm
U\right)  =\left(  z\frac{d}{dz}+\nu\right)  ^{2}-U\left(  z\right)  ^{2}\pm
z\frac{d}{dz}U\left(  z\right) \\
U\left(  z\right)  \equiv\sum_{k>0}\upsilon_{k}z^{k}%
\end{gather}
where the exponents $k$\ in the \textquotedblleft
superpotential\textquotedblright\ $U$ are of the \emph{same} sign, and where
$\nu$ and $\upsilon_{k}$ have arbitrary values. \ We look for eigenfunctions
and associated functions that have particular analytic behavior near $z=0$.
\ The connection to PT symmetric theories is achieved by writing $z=me^{ix}$,
as explained in \cite{CM}. \ 

\section{General Theory for $\nu=0$}

First we consider $\nu=0$. \ For bounded functions of $x\in\left(
-\infty,+\infty\right)  $, with $z=me^{ix}$, the spectrum of any Hamiltonian
of the form \cite{NonSelfAdjointMathWork,CM}%
\begin{equation}
H=\left(  z\frac{d}{dz}\right)  ^{2}+\sum_{k>0}\mu_{k}z^{k}%
\end{equation}
is known to be the real, positive half-line, for \emph{any} choice of $\mu$s
such that $\sum_{k>0}\left\vert \mu_{k}\right\vert <\infty$, and not just for
those $\mu$s which are real or $x$-translationally equivalent to real values.
\ So PT symmetry is not required for real energy eigenvalues in these models
\cite{Mostafazadeh}. \ Here, we will restrict our attention to $2\pi$-periodic
functions of $x$, with $z=me^{ix}$, and their duals, to obtain a discrete
subset of real energy eigenvalues, namely just $\left\{  E_{n}=n^{2}%
\ |\ n=0,1,\cdots\right\}  $. \ 

In this situation the general theory for supersymmetric pairs of non-hermitian
Hamiltonians goes as follows. \ For a given $U\left(  z\right)  $ we may
construct pairs of \emph{finite polynomials}\ in $z^{-1}$,%
\begin{equation}
\chi_{n}^{\pm}\left(  z\right)  =\frac{1}{z^{n}}\sum_{j=0}^{n}c_{n,j}^{\pm
}z^{j} \label{DualPolys}%
\end{equation}
that satisfy the inhomogeneous equations,
\begin{equation}
\left(  z\frac{d}{dz}\pm U\left(  z\right)  \right)  \chi_{n}^{\pm}\left(
z\right)  +n\chi_{n}^{\mp}\left(  z\right)  =\Lambda_{n}^{\pm}\left(
z\right)  \equiv\sum_{k>0}\lambda_{n}^{\pm}\left(  k\right)  z^{k}
\label{DualPolyEquations}%
\end{equation}
where $U\left(  z\right)  $ and $\Lambda_{n}^{\pm}\left(  z\right)  $ are
analytic about the origin, at which point they all vanish. \ A priori the
$\Lambda_{n}^{\pm}\left(  z\right)  $ need not be given, for a given $U\left(
z\right)  $ they may be determined along with $\chi_{n}^{\pm}\left(  z\right)
$. \ Clearly, $\chi_{n}^{\pm}\left(  z\right)  \ _{\widetilde{z\rightarrow0}%
}\ c_{n,0}^{\pm}/z^{n}$ and, by convention, we normalize so that the
coefficient of the most negative power of $z$ is just $c_{n,0}^{\pm}=1$.
\ This choice also fixes the normalization of $\Lambda_{n}^{\pm}$. \ For a
given $U\left(  z\right)  $ the four functions $\chi_{n}^{\pm}\left(
z\right)  $ and $\Lambda_{n}^{\pm}\left(  z\right)  $\ are now completely specified.

Alternatively, we construct pairs of functions $\psi_{n}^{\pm}\left(
z\right)  $\ analytic about the origin, hence given by series in non-negative
integer powers of $z$, that satisfy the equations
\begin{equation}
\left(  z\frac{d}{dz}\pm U\left(  z\right)  \right)  \psi_{n}^{\pm}\left(
z\right)  =n\psi_{n}^{\mp}\left(  z\right)  \label{EigenEquation}%
\end{equation}
with $\psi_{n}^{\pm}\left(  z\right)  \ _{\widetilde{z\rightarrow0}}\ z^{n}$.
\ Usually these are infinite series, but again, for a given $U\left(
z\right)  $, both functions $\psi_{n}^{\pm}\left(  z\right)  $\ are now
completely specified.

It follows that the $\psi_{n}^{\pm}\left(  z\right)  $ are eigenfunctions of
$H_{\pm}$ with eigenvalues $n^{2}$,
\begin{gather}
H_{\pm}\psi_{n}^{\pm}\left(  z\right)  =n^{2}\psi_{n}^{\pm}\left(  z\right)
\label{IntegerSquaredEnergyEqn}\\
H_{\pm}=\left(  z\frac{d}{dz}\mp U\left(  z\right)  \right)  \left(  z\frac
{d}{dz}\pm U\left(  z\right)  \right)  =\left(  z\frac{d}{dz}\right)
^{2}-U\left(  z\right)  ^{2}\pm z\frac{d}{dz}U\left(  z\right)
\end{gather}
while the associated polynomials obey the inhomogeneous equations%
\begin{equation}
\left(  H_{\pm}-k^{2}\right)  \chi_{k}^{\pm}\left(  z\right)  =\left(
z\frac{d}{dz}\mp U\left(  z\right)  \right)  \Lambda_{k}^{\pm}\left(
z\right)  -k\Lambda_{k}^{\mp}\left(  z\right)
\label{IntegerSquaredDualPolyEqn}%
\end{equation}
Note the RHS of this last equation involves only positive powers, $z^{n}$ for
$n\geq1$. \ Moreover, it also follows that these functions are biorthonormal
systems such that
\begin{equation}
\frac{1}{2\pi i}\oint\frac{dz}{z}\ \chi_{k}^{\pm}\left(  z\right)  \psi
_{n}^{\pm}\left(  z\right)  =\delta_{kn}\ ,\ \ \ \sum_{n=0}^{\infty}\chi
_{n}^{\pm}\left(  w\right)  \psi_{n}^{\pm}\left(  z\right)  =\frac{1}%
{1-\frac{z}{w}} \label{OrthoComplete}%
\end{equation}
These relations encode orthonormality and completeness on analytic functions.
\ For more or less obvious reasons, we will call the full set of functions
$\left\{  \chi_{j}^{+},\psi_{k}^{+};\chi_{l}^{-},\psi_{n}^{-}\right\}  $ an
analytic, \emph{supersymmetric biorthogonal system} corresponding to the given
function $U\left(  z\right)  $. \ 

\paragraph{The dual polynomials and their inhomogeneities.}

We start the construction with the dual polynomials\footnote{So these systems
are exceptions to the statement: \ \textquotedblleft A direct and frontal
attack on the problem of determining [the dual polynomials] is not usually
fruitful.\textquotedblright\ -- Morse and Feshbach \cite{BiorthogonalClassics}
p 931.} and explicitly compute the first few. \ This serves to illustrate how
the various functions are uniquely determined, given that\ $\chi_{n}^{\pm
}\left(  z\right)  \ _{\widetilde{z\rightarrow0}}\ 1/z^{n}$. \ We choose the
coefficients in $\chi_{n}^{\pm}\left(  z\right)  $ to eliminate all $z^{-k}$
terms on the LHS of (\ref{DualPolyEquations}), for $k=0,1,\cdots,n$, and then
we sweep all the remaining positive powers of $z$ into the $\Lambda_{n}^{\pm
}\left(  z\right)  $. \ We find

\begin{center}
$\boldsymbol{\chi}$\textbf{\ and }$\boldsymbol{\Lambda}$ \textbf{Table}
\end{center}

\noindent\hspace{-0.25in}$%
\begin{array}
[c]{cc}%
\chi_{0}^{\pm}\left(  z\right)  \medskip=1 & \Lambda_{0}^{\pm}\left(
z\right)  =\pm U\left(  z\right)  =\pm\sum_{k>0}\upsilon_{k}z^{k}\\
\chi_{1}^{\pm}\left(  z\right)  \medskip=\dfrac{1}{z}\pm\upsilon_{1} &
\Lambda_{1}^{\pm}\left(  z\right)  =\sum_{k>0}\left(  \upsilon_{1}\upsilon
_{k}\pm\upsilon_{k+1}\right)  z^{k}\\
\chi_{2}^{\pm}\left(  z\right)  \medskip=\dfrac{1}{z^{2}}\pm\frac{1}%
{3}\upsilon_{1}\dfrac{1}{z}\pm\frac{1}{2}\upsilon_{2}-\frac{1}{6}\upsilon
_{1}^{2} & \Lambda_{2}^{\pm}\left(  z\right)  =\sum_{k>0}\left(  \left(
\frac{1}{2}\upsilon_{2}\mp\frac{1}{6}\upsilon_{1}^{2}\right)  \upsilon
_{k}+\frac{1}{3}\upsilon_{1}\upsilon_{k+1}\pm\upsilon_{k+2}\right)  z^{k}\\
\chi_{3}^{\pm}\left(  z\right)  =\left(
\begin{array}
[c]{c}%
\dfrac{1}{z^{3}}\pm\frac{1}{5}\upsilon_{1}\dfrac{1}{z^{2}}+\left(  \pm\frac
{1}{4}\upsilon_{2}-\frac{1}{10}\upsilon_{1}^{2}\right)  \dfrac{1}{z}\medskip\\
\pm\frac{1}{3}\upsilon_{3}-\frac{3}{20}\upsilon_{2}\upsilon_{1}\mp\frac{1}%
{30}\upsilon_{1}^{3}\medskip
\end{array}
\right)  & \Lambda_{3}^{\pm}\left(  z\right)  =\sum_{k>0}\left(
\begin{array}
[c]{c}%
\left(  \frac{1}{3}\upsilon_{3}\mp\frac{3}{20}\upsilon_{2}\upsilon_{1}%
-\frac{1}{30}\upsilon_{1}^{3}\right)  \upsilon_{k}\medskip\\
+\left(  \frac{1}{4}\upsilon_{2}\mp\frac{1}{10}\upsilon_{1}^{2}\right)
\upsilon_{k+1}+\frac{1}{5}\upsilon_{1}\upsilon_{k+2}\pm\upsilon_{k+3}\medskip
\end{array}
\right)  z^{k}%
\end{array}
\medskip$

\noindent etc. \ \ When the sum $\sum_{k>0}$\ that defines $U$\ is finite, the
process is clearly finite mathematics all the way, for any $n$. \ Note that
$\chi_{n}^{-}\left(  z\right)  $ and $\Lambda_{n}^{-}\left(  z\right)  $ are
obtained from $\chi_{n}^{+}\left(  z\right)  $ and $\Lambda_{n}^{+}\left(
z\right)  $, or vice versa, just by flipping the signs of all the $\upsilon$s.
\ Also note that all negative powers of $z$ can be expressed as finite sums of
\emph{either} $\left\{  \chi_{n}^{+}\left(  z\right)  \ |\ n\geq0\right\}  $
\emph{or} $\left\{  \chi_{n}^{-}\left(  z\right)  \ |\ n\geq0\right\}  $.

More systematically, we solve (\ref{DualPolyEquations}) as follows. \ We
impose the condition that the RHS involve only positive powers of $z$ so as to
obtain an inhomogeneity that will be orthogonal to the span of $\left\{
z^{n}\ |\ n\geq0\right\}  $ under contour integration $%
{\textstyle\oint}
\frac{dz}{z}\cdots$. \ This leads to $n$ equations that fix the coefficients
$c_{n,k}$ for $k=1,\cdots,n$ in terms of $c_{n,0}$, the latter being an
overall choice of normalization. \ Thus%
\begin{gather}
\sum_{j=0}^{n}c_{n,j}^{\pm}\left(  j-n\right)  z^{j}\pm\sum_{k>0}\upsilon
_{k}z^{k}\sum_{j=0}^{n}c_{n,j}^{\pm}z^{j}+n\sum_{j=0}^{n}c_{n,j}^{\mp}%
z^{j}=\sum_{k>0}\lambda_{n}^{\pm}\left(  k\right)  z^{k+n}\\
\text{so \ \ }c_{n,j}^{\pm}\left(  j-n\right)  \pm\sum_{k=1}^{j}\upsilon
_{k}c_{n,j-k}^{\pm}+nc_{n,j}^{\mp}=0\text{ \ \ for \ \ }j=0,\cdots,n\\
\text{and \ \ }\lambda_{n}^{\pm}\left(  k\right)  =\pm\sum_{j=0}^{n}%
\upsilon_{k+n-j}c_{n,j}^{\pm} \label{InhomogeneityCoefficients}%
\end{gather}
We determine all the coefficients in the $\chi_{n}^{\pm}\left(  z\right)  $
from the pair of equations%
\[
\left(
\begin{array}
[c]{cc}%
n-j & -n\\
n & j-n
\end{array}
\right)  \left(
\begin{array}
[c]{c}%
c_{n,j}^{+}\\
c_{n,j}^{-}%
\end{array}
\right)  =\sum_{k=1}^{j}\upsilon_{k}\left(
\begin{array}
[c]{c}%
c_{n,j-k}^{+}\\
c_{n,j-k}^{-}%
\end{array}
\right)  \text{ \ \ for \ \ }j=0,\cdots,n
\]
where $c_{n,0}^{\pm}\equiv1$. \ Now $\det\left(
\begin{array}
[c]{cc}%
n-j & -n\\
n & j-n
\end{array}
\right)  =j\left(  2n-j\right)  >0$ for $0<j\leq n$, and\ $\left(
\begin{array}
[c]{cc}%
n-j & -n\\
n & j-n
\end{array}
\right)  ^{-1}=\frac{1}{j\left(  2n-j\right)  }\left(
\begin{array}
[c]{cc}%
j-n & n\\
-n & n-j
\end{array}
\right)  $. \ So we have the recursion relations%
\begin{equation}
c_{n,j}^{\pm}=\frac{\pm1}{j\left(  2n-j\right)  }\sum_{k=1}^{j}\upsilon
_{k}\left(  \left(  j-n\right)  c_{n,j-k}^{\pm}+nc_{n,j-k}^{\mp}\right)
\label{cRecursion}%
\end{equation}
Each $c_{n,j}^{\pm}$ depends on only the first $j$ coefficients in the
expansion for $U\left(  z\right)  $, i.e. just on $\upsilon_{k\leq j}$. \ From
the $c_{n,j}^{\pm}$\ we then determine the $\lambda_{n}^{\pm}\left(  k\right)
$\ using (\ref{InhomogeneityCoefficients}). \ Note that $\lambda_{n}^{\pm
}\left(  k\right)  \neq0$ for $k>n$ is \emph{possible} here, depending on the
values of the $\upsilon$s. \ The $\lambda_{n}^{\pm}\left(  k\right)  $ will
depend on all the $\upsilon$s, in general, with $k$ taking on all values up to
and including the highest power of $z$ appearing in $U$.

For example, for $n=2$:%
\begin{align}
c_{2,0}^{\pm}  &  =1\ ,\ \ \ c_{2,1}^{\pm}=\pm\frac{1}{3}\upsilon
_{1}\ ,\ \ \ c_{2,2}^{\pm}=\pm\frac{1}{2}\upsilon_{2}-\frac{1}{6}\upsilon
_{1}^{2}\\
\pm\lambda_{2}^{\pm}\left(  k\right)   &  =\upsilon_{k+2}+\upsilon
_{k+1}c_{2,1}^{\pm}+\upsilon_{k}c_{2,2}^{\pm}\nonumber\\
&  =\upsilon_{k+2}\pm\frac{1}{3}\upsilon_{1}\upsilon_{k+1}+\left(  \pm\frac
{1}{2}\upsilon_{2}-\frac{1}{6}\upsilon_{1}^{2}\right)  \upsilon_{k}%
\end{align}
And for $n=3$:%
\begin{align}
c_{3,0}^{\pm}  &  =1\ ,\ \ \ c_{3,1}^{\pm}=\pm\frac{1}{5}\upsilon
_{1}\ ,\ \ \ c_{3,2}^{\pm}=\pm\frac{1}{4}\upsilon_{2}-\frac{1}{10}\upsilon
_{1}^{2}\ ,\ \ \ c_{3,3}^{\pm}=\pm\frac{1}{3}\upsilon_{3}-\frac{3}{20}%
\upsilon_{2}\upsilon_{1}\mp\frac{1}{30}\upsilon_{1}^{3}\\
\pm\lambda_{3}^{\pm}\left(  k\right)   &  =\upsilon_{k+3}+\upsilon
_{k+2}c_{3,1}^{\pm}+\upsilon_{k+1}c_{3,2}^{\pm}+\upsilon_{k}c_{3,3}^{\pm
}\nonumber\\
&  =\upsilon_{k+3}\pm\frac{1}{5}\upsilon_{1}\upsilon_{k+2}+\left(  \pm\frac
{1}{4}\upsilon_{2}-\frac{1}{10}\upsilon_{1}^{2}\right)  \upsilon_{k+1}+\left(
\pm\frac{1}{3}\upsilon_{3}-\frac{3}{20}\upsilon_{2}\upsilon_{1}\mp\frac{1}%
{30}\upsilon_{1}^{3}\right)  \upsilon_{k}%
\end{align}
Hence the table given earlier.

\paragraph{Energy eigenfunctions}

The energy eigenfunctions, when written as series,
\begin{equation}
\psi_{n}^{\pm}\left(  z\right)  =z^{n}\sum_{j=0}^{\infty}a_{n,j}^{\pm}z^{j}%
\end{equation}
can be determined just as the dual polynomials were by direct solution of
(\ref{EigenEquation}), or else the eigenfunctions can be determined by
imposing the bi-orthonormalizations in (\ref{OrthoComplete}). As discussed in
\cite{CM}, these orthogonality conditions amount to a set of triangular
equations which can always be solved, sequentially, for the $a_{n,j}^{\pm}$ in
terms of the $c^{\pm}$s. \ Namely%
\begin{equation}
a_{n,0}^{\pm}=\frac{1}{c_{n,0}^{\pm}}\ ,\ \ \ \sum_{j=0}^{k-n}c_{k,k-n-j}%
^{\pm}a_{n,j}^{\pm}=0\text{ \ \ for \ \ }k>n \label{Recursions}%
\end{equation}
The series for $\psi_{n}^{\pm}$\ is a development in the minors that invert
these triangular equations. \ By considering all $k>n$ in succession, we
obtain all $a_{n,j}^{\pm}$ in terms of $c_{k,l}^{\pm}$, or vice versa. \ 

For convenience, we again choose the normalizations $c_{n,0}^{\pm}=1$. \ The
results of the recursion relations (\ref{Recursions}) are the pair of
correlated series%
\begin{align}
\chi_{n}^{\pm}\left(  z\right)   &  =\frac{1}{z^{n}}\left(  1+c_{n,1}^{\pm
}z+c_{n,2}^{\pm}z^{2}+\cdots+c_{n,n}^{\pm}z^{n}\right) \label{ChiSeries}\\
\psi_{n}^{\pm}\left(  z\right)   &  =z^{n}\left\vert
\begin{array}
[c]{ccccc}%
1 & z & z^{2} & z^{3} & \cdots\\
c_{n+1,1}^{\pm} & 1 & 0 & 0 & \cdots\\
c_{n+2,2}^{\pm} & c_{n+2,1}^{\pm} & 1 & 0 & \cdots\\
c_{n+3,3}^{\pm} & c_{n+3,2}^{\pm} & c_{n+3,1}^{\pm} & 1 & \cdots\\
\vdots & \vdots & \vdots & \vdots & \ddots
\end{array}
\right\vert \label{PsiSeries}\\
&  =z^{n}\left(  1-c_{n+1,1}^{\pm}z+\left\vert
\begin{array}
[c]{cc}%
c_{n+1,1}^{\pm} & 1\\
c_{n+2,2}^{\pm} & c_{n+2,1}^{\pm}%
\end{array}
\right\vert z^{2}-\left\vert
\begin{array}
[c]{ccc}%
c_{n+1,1}^{\pm} & 1 & 0\\
c_{n+2,2}^{\pm} & c_{n+2,1}^{\pm} & 1\\
c_{n+3,3}^{\pm} & c_{n+3,2}^{\pm} & c_{n+3,1}^{\pm}%
\end{array}
\right\vert z^{3}+-\cdots\right) \nonumber
\end{align}
Alternatively the $c^{\pm}$s may be expressed in terms of the $a^{\pm}$s.%
\begin{align}
\psi_{n}^{\pm}\left(  z\right)   &  =z^{n}\left(  1+a_{n,1}^{\pm}%
z+a_{n,2}^{\pm}z^{2}+a_{n,3}^{\pm}z^{3}+a_{n,4}^{\pm}z^{4}+\cdots\right)
\label{OtherPsiSeries}\\
\chi_{n}^{\pm}\left(  z\right)   &  =\frac{1}{z^{n}}\left\vert
\begin{array}
[c]{ccccccc}%
1 & z & z^{2} & z^{3} & \cdots & z^{n-1} & z^{n}\\
a_{n-1,1}^{\pm} & 1 & 0 & 0 & \cdots & 0 & 0\\
a_{n-2,2}^{\pm} & a_{n-2,1}^{\pm} & 1 & 0 & \cdots & 0 & 0\\
a_{n-3,3}^{\pm} & a_{n-3,2}^{\pm} & a_{n-3,1}^{\pm} & 1 & \cdots & 0 & 0\\
\vdots & \vdots & \vdots & \vdots & \ddots & \vdots & \vdots\\
a_{1,n-1}^{\pm} & a_{1,n-2}^{\pm} & a_{1,n-3}^{\pm} & a_{1,n-4}^{\pm} & \cdots
& 1 & 0\\
a_{0,n}^{\pm} & a_{0,n-1}^{\pm} & a_{0,n-2}^{\pm} & a_{0,n-3}^{\pm} & \cdots &
a_{0,1}^{\pm} & 1
\end{array}
\right\vert \label{OtherChiSeries}\\
&  =\dfrac{1}{z^{n}}\left(  1-a_{n-1,1}^{\pm}z+\left\vert
\begin{array}
[c]{cc}%
a_{n-1,1}^{\pm} & 1\\
a_{n-2,2}^{\pm} & a_{n-2,1}^{\pm}%
\end{array}
\right\vert z^{2}-\left\vert
\begin{array}
[c]{ccc}%
a_{n-1,1}^{\pm} & 1 & 0\\
a_{n-2,2}^{\pm} & a_{n-2,1}^{\pm} & 1\\
a_{n-3,3}^{\pm} & a_{n-3,2}^{\pm} & a_{n-3,1}^{\pm}%
\end{array}
\right\vert z^{3}\right. \nonumber\\
&  +-\cdots+\left(  -1\right)  ^{n}\left\vert
\begin{array}
[c]{ccccc}%
a_{n-1,1}^{\pm} & 1 & 0 & \cdots & 0\\
a_{n-2,2}^{\pm} & a_{n-2,1}^{\pm} & 1 & \cdots & 0\\
a_{n-3,3}^{\pm} & a_{n-3,2}^{\pm} & a_{n-3,1}^{\pm} & 1 & \vdots\\
\vdots & \vdots & \vdots & \ddots & 1\\
a_{0,n}^{\pm} & a_{0,n-1}^{\pm} & a_{0,n-2}^{\pm} & \cdots & a_{0,1}^{\pm}%
\end{array}
\right\vert z^{n}\left.
\begin{array}
[c]{c}%
\ ^{\ }\\
\ \\
\ ^{\ }%
\end{array}
\right) \nonumber
\end{align}

\paragraph{Orthogonality and Completeness}

With either pair of these correlated series, the orthogonality relations in
(\ref{OrthoComplete})\ are easily checked. \ They amount to the obvious
statements that
\begin{equation}
\frac{1}{2\pi i}\oint\frac{dz}{z}\ \chi_{n}^{\pm}\left(  z\right)  \psi
_{n}^{\pm}\left(  z\right)  =1\ ,\ \ \ \frac{1}{2\pi i}\oint\frac{dz}{z}%
\ \chi_{n}^{\pm}\left(  z\right)  \psi_{n+k}^{\pm}\left(  z\right)  =0\text{
\ \ for \ \ }k>0
\end{equation}
as well as more involved cancellations to show $\frac{1}{2\pi i}\oint\frac
{dz}{z}\ \chi_{n+k}^{\pm}\left(  z\right)  \psi_{n}^{\pm}\left(  z\right)  =0$
for $k>0$. \ In particular, there is a complete cancellation of all the
contributions to this latter contour integral upon expansion of the $\left(
k+1\right)  \times\left(  k+1\right)  $ determinant that appears in the first
line of (\ref{OtherChiSeries}). \ Exploiting the multi-linearity of the
determinant, and performing the integrations entry by entry,%
\begin{gather}
\frac{1}{2\pi i}\oint\frac{dz}{z}\ \chi_{n+k}^{\pm}\left(  z\right)  \psi
_{n}^{\pm}\left(  z\right)  =\frac{1}{2\pi i}\oint\frac{dz}{z}\ z^{-k}\left(
1+a_{n,1}^{\pm}z+a_{n,2}^{\pm}z^{2}+a_{n,3}^{\pm}z^{3}+a_{n,4}^{\pm}%
z^{4}+\cdots\right)  \times\nonumber\\
\times\left\vert
\begin{array}
[c]{ccccccc}%
1 & z & z^{2} & z^{3} & \cdots & z^{n+k-1} & z^{n+k}\\
a_{n+k-1,1}^{\pm} & 1 & 0 & 0 & \cdots & 0 & 0\\
a_{n+k-2,2}^{\pm} & a_{n+k-2,1}^{\pm} & 1 & 0 & \cdots & 0 & 0\\
a_{n+k-3,3}^{\pm} & a_{n+k-3,2}^{\pm} & a_{n+k-3,1}^{\pm} & 1 & \cdots & 0 &
0\\
\vdots & \vdots & \vdots & \vdots & \ddots & \vdots & \vdots\\
a_{1,n+k-1}^{\pm} & a_{1,n+k-2}^{\pm} & a_{1,n+k-3}^{\pm} & a_{1,n+k-4}^{\pm}
& \cdots & 1 & 0\\
a_{0,n+k}^{\pm} & a_{0,n+k-1}^{\pm} & a_{0,n+k-2}^{\pm} & a_{0,n+k-3}^{\pm} &
\cdots & a_{0,1}^{\pm} & 1
\end{array}
\right\vert \nonumber\\
\nonumber\\
=\left\vert
\begin{array}
[c]{cccccccc}%
a_{n,k}^{\pm} & a_{n,k-1}^{\pm} & a_{n,k-2}^{\pm} & a_{n,k-3}^{\pm} & \cdots &
a_{n,0}^{\pm}=1 & 0 & \cdots\\
a_{n+k-1,1}^{\pm} & 1 & 0 & 0 & \cdots & 0 & 0 & \cdots\\
a_{n+k-2,2}^{\pm} & a_{n+k-2,1}^{\pm} & 1 & 0 & \cdots & 0 & 0 & \cdots\\
a_{n+k-3,3}^{\pm} & a_{n+k-3,2}^{\pm} & a_{n+k-3,1}^{\pm} & 1 & \cdots & 0 &
0 & \cdots\\
\vdots & \vdots & \vdots & \vdots & \ddots & \vdots & \vdots & \vdots\\
a_{n,k}^{\pm} & a_{n,k-1}^{\pm} & a_{n,k-2}^{\pm} & a_{n,k-3}^{\pm} & \cdots &
1 & 0 & \cdots\\
a_{n-1,k+1}^{\pm} & a_{n-1,k}^{\pm} & a_{n-1,k-1}^{\pm} & a_{n-1,k-2}^{\pm} &
\cdots & a_{0,1}^{\pm} & 1 & \cdots\\
\vdots & \vdots & \vdots & \vdots & \cdots & \vdots & \vdots & \ddots
\end{array}
\right\vert
\end{gather}
which vanishes since the $1$st and the $\left(  k+1\right)  $th rows are
identical. \ Similarly%
\begin{equation}
\frac{1}{2\pi i}\oint\frac{dz}{z}\ \chi_{n+k}^{\pm}\left(  z\right)  \psi
_{n}^{\pm}\left(  z\right)  =\left\vert
\begin{array}
[c]{ccccc}%
c_{n+k,k}^{\pm} & c_{n+k,k-1}^{\pm} & c_{n+k,k-2}^{\pm} & \cdots & 1\\
c_{n+1,1}^{\pm} & 1 & 0 & \cdots & 0\\
c_{n+2,2}^{\pm} & c_{n+2,1}^{\pm} & 1 & \cdots & 0\\
\vdots & \vdots & \vdots & \ddots & \vdots\\
c_{n+k,k}^{\pm} & c_{n+k,k-1}^{\pm} & c_{n+k,k-2}^{\pm} & \cdots & 1
\end{array}
\right\vert =0
\end{equation}
for $k>0$.

The correlations between coefficients in (\ref{ChiSeries}) and
(\ref{PsiSeries}), or in (\ref{OtherPsiSeries}) and (\ref{OtherChiSeries}),
also imply the completeness relation in (\ref{OrthoComplete}) by guaranteeing
that all terms of the form $z^{k}w^{-l}$ for $k\neq l$ cancel out in the sum
$\sum_{n=0}^{\infty}\chi_{n}^{\pm}\left(  w\right)  \psi_{n}^{\pm}\left(
z\right)  $. \ This cancellation is encoded in the identities%
\begin{equation}
0=\sum_{n=0}^{\infty}\sum_{l=1}^{\infty}a_{n,l}^{\pm}w^{n+l}z^{-n}+\sum
_{n=1}^{\infty}\sum_{k=1}^{n}c_{n,k}^{\pm}w^{n}z^{k-n}+\sum_{n=1}^{\infty}%
\sum_{l=1}^{\infty}\sum_{k=1}^{n}a_{n,l}^{\pm}c_{n,k}^{\pm}w^{n+l}z^{k-n}%
\end{equation}
which follow from (\ref{Recursions}). \ Terms of the form $\left(  z/w\right)
^{k}$\ as required to give $\frac{1}{1-\frac{z}{w}}$ are provided just by the
leading terms in (\ref{ChiSeries}) and (\ref{PsiSeries}).

The coefficients of $z^{n+k}$\ in $\psi_{n}^{\pm}\left(  z\right)  $\ are
again finite polynomials in the $\upsilon$s. \ While convergence of this
series, as written, is certainly not obvious for arbitrary $\upsilon$s, it is
clear that convergence can be determined on a case-by-case basis from the
explicit form of the coefficients. \ Moreover, when the \emph{number} of
$\upsilon$s is finite, no matter what their values are, it is not too
difficult to show the $\psi_{n}^{\pm}\left(  z\right)  $ are entire functions
of $z$. \ Thus, if we assume the requisite convergence of the $\psi_{n}^{\pm}$
series, all the $\left\{  \psi_{n}^{\pm}\left(  z\right)  \ |\text{ }%
n\geq0\right\}  $ are determined, and either $\left\{  \psi_{n}^{+}\left(
z\right)  \ |\text{ }n\geq0\right\}  $\ or $\left\{  \psi_{n}^{-}\left(
z\right)  \ |\ n\geq0\right\}  $ are complete on the span of $\left\{
z^{n}\ |\ n\geq0\right\}  $. \ All positive powers of $z$ can be expressed as
series of $\left\{  \psi_{n}^{\pm}\left(  z\right)  \right\}  $, just as all
negative powers of $z$ can be expressed as finite sums of $\left\{  \chi
_{n}^{\pm}\left(  z\right)  \right\}  $.

Remarkably, the non-degenerate energy eigenfunctions $\left\{  \psi_{n}^{\pm
}\left(  z\right)  \right\}  $ just obtained turn out to be \emph{all} of the
eigenfunctions of $H^{\pm}$ which are $2\pi$-periodic in $x$, where
$z=me^{ix}$. \ Moreover, the fact that $\psi_{n}^{\pm}\left(  z\right)  $\ are
indeed eigenfunctions, as given in (\ref{IntegerSquaredEnergyEqn}),\ can be
deduced in a novel way from (\ref{IntegerSquaredDualPolyEqn}), the
biorthogonality of $\left\{  \chi_{j}^{\pm}\left(  z\right)  ,\psi_{k}^{\pm
}\left(  z\right)  \right\}  $, and the completeness of $\left\{  \psi
_{n}^{\pm}\left(  z\right)  \right\}  $ for analytic functions about the
origin, as described in \cite{CM}. \ In fact, the argument given in \cite{CM}
can be adapted to the first-order equations. \ Completeness on analytic
functions about $z=0$\ allows us to write
\begin{equation}
\left(  z\frac{d}{dz}\pm U\left(  z\right)  \right)  \psi_{n}^{\pm}\left(
z\right)  =\sum_{k=n}^{\infty}b_{n,k}^{\pm}\psi_{k}^{\mp}\left(  z\right)  .
\end{equation}
Note the chosen interchange $\psi^{\pm}\leftrightarrow\psi^{\mp}$ upon
LHS$\leftrightarrow$RHS. \ From this expansion and biorthonormality, we have
$b_{n,k}^{\pm}=\frac{1}{2\pi i}%
{\textstyle\oint}
\frac{dz}{z}\chi_{k}^{\mp}\left(  z\right)  \left(  z\frac{d}{dz}\pm U\left(
z\right)  \right)  \psi_{n}^{\pm}\left(  z\right)  $. $\ $But then, upon
integrating by parts and using (\ref{DualPolyEquations}) as well as the
orthonormality relations, we also have $\frac{1}{2\pi i}%
{\textstyle\oint}
\frac{dz}{z}\chi_{k}^{\mp}\left(  z\right)  \left(  z\frac{d}{dz}\pm U\left(
z\right)  \right)  \psi_{n}^{\pm}\left(  z\right)  =n\delta_{k,n}$. \ So
$b_{n,k}=n\delta_{k,n}$, and (\ref{EigenEquation}) is obtained.\footnote{Note
that $z=0$ is a regular singular point of (\ref{EigenEquation}) and/or
(\ref{IntegerSquaredEnergyEqn}), for any number and any values of the
$\upsilon$s such that $\sum_{k>0}\upsilon_{k}z^{k}$ converges near the origin.
\ In fact, (\ref{PsiSeries}) is exactly the conventional series obtained by
expanding about the regular singular point at $z=0$, albeit the series was
obtained here in an unusual way from the properties of the dual space
polynomials.} \ Conversely, given (\ref{EigenEquation}) and
(\ref{DualPolyEquations}), we may prove the orthogonality relations $%
{\textstyle\oint}
\frac{dz}{z}\chi_{k}^{\pm}\left(  z\right)  \psi_{n}^{\pm}\left(  z\right)
=0$ for $k^{2}\neq n^{2}\neq0$ just by inserting $\left(  z\frac{d}{dz}\pm
U\left(  z\right)  \right)  $ and integrating by parts. \ That is to say%
\begin{align}
n%
{\textstyle\oint}
\frac{dz}{z}\chi_{k}^{\pm}\left(  z\right)  \psi_{n}^{\pm}\left(  z\right)
&  =%
{\textstyle\oint}
\frac{dz}{z}\chi_{k}^{\pm}\left(  z\right)  \left(  z\frac{d}{dz}\mp U\left(
z\right)  \right)  \psi_{n}^{\mp}\left(  z\right)  =%
{\textstyle\oint}
\frac{dz}{z}\psi_{n}^{\mp}\left(  z\right)  \left(  -z\frac{d}{dz}\mp U\left(
z\right)  \right)  \chi_{k}^{\pm}\left(  z\right) \nonumber\\
&  =k%
{\textstyle\oint}
\frac{dz}{z}\chi_{k}^{\mp}\left(  z\right)  \psi_{n}^{\mp}\left(  z\right)
\end{align}
Thus $\left(  n^{2}-k^{2}\right)
{\textstyle\oint}
\frac{dz}{z}\chi_{k}^{\pm}\left(  z\right)  \psi_{n}^{\pm}\left(  z\right)
=0$.

\section{Examples}

As an explicit example, to parallel the discussion in \cite{CM}, we note that
the superpotential\footnote{Actually, for any constant $c$ the superpotential
$U\left(  z\right)  =z\times\frac{J_{1}\left(  z\right)  +cY_{1}\left(
z\right)  }{J_{0}\left(  z\right)  +cY_{0}\left(  z\right)  }$ gives a simple
quadratic potential for $H_{+}$, $z^{2}=-U^{2}+z\frac{d}{dz}U$. \ But if
$c\neq0$, $U$ involves a logarithm, $\ln z$, and hence is not a periodic
function of $x$ for $z=me^{ix}$.}%
\begin{equation}
U\left(  z\right)  =\frac{zJ_{1}\left(  z\right)  }{J_{0}\left(  z\right)  }%
\end{equation}
gives a simple quadratic potential for $H_{+}$%
\begin{equation}
V_{+}=-U^{2}+z\frac{d}{dz}U=z^{2}%
\end{equation}
but a much more complicated partner potential for $H_{-}$%
\begin{align}
V_{-}  &  =-U^{2}-z\frac{d}{dz}U=-z^{2}\left(  1+2\left(  \frac{J_{1}\left(
z\right)  }{J_{0}\left(  z\right)  }\right)  ^{2}\right) \\
&  =-z^{2}-\frac{1}{2}z^{4}-\frac{1}{8}z^{6}-\frac{11}{384}z^{8}+O\left(
z^{10}\right)
\end{align}
The complexity of $H_{-}$ suggests that we seek a simpler $U$ to fully
illustrate the general theory.

\paragraph{Complex Morse potentials}

Again referring to \cite{CM}, we\ consider $U\left(  z\right)  =\mu z$, hence
\begin{equation}
H_{\pm}=\left(  z\frac{d}{dz}+\nu\right)  ^{2}\pm\mu z-\mu^{2}z^{2}%
\end{equation}
Note that for this simple example the solution of one Hamiltonian, say $H_{+}
$, immediately gives the solution for the other, through the relations
$H_{-}\left[  z\right]  =H_{+}\left[  -z\right]  $, $\psi_{n}^{-}\left(
z\right)  \propto\psi_{n}^{+}\left(  -z\right)  $. \ But this is not
necessarily the most transparent way to write the solutions for $H_{-}$. \ 

When the vector potential is not present, $\nu=0$, it may be best to simply
note%
\begin{equation}
\left(  z\frac{d}{dz}\pm z\right)  \left(  \sqrt{\frac{z}{2}}\left(
I_{n-1/2}\left(  z\right)  \mp I_{n+1/2}\left(  z\right)  \right)  \right)
=n\left(  \sqrt{\frac{z}{2}}\left(  I_{n-1/2}\left(  z\right)  \pm
I_{n+1/2}\left(  z\right)  \right)  \right)
\end{equation}
So then it is obvious that
\begin{equation}
\psi_{n}^{\pm}\left(  z\right)  =Z_{n}^{\pm}\sqrt{\frac{z}{2}}\left(
I_{n-1/2}\left(  z\right)  \mp I_{n+1/2}\left(  z\right)  \right)
\end{equation}
are eigenfunctions of%
\begin{equation}
H_{\pm}=\left(  z\frac{d}{dz}\right)  ^{2}\pm z-z^{2}=\left(  z\frac{d}{dz}\mp
z\right)  \left(  z\frac{d}{dz}\pm z\right)
\end{equation}
with eigenvalues $n^{2}$\ as given by%
\begin{equation}
H_{\pm}\psi_{n}^{\pm}\left(  z\right)  =n^{2}\psi_{n}^{\pm}\left(  z\right)
\end{equation}
and with normalization constants $Z_{n}^{\pm}$. \ Other ways to write the
eigenfunctions for this example are:%
\begin{align}
\psi_{n}^{\pm}\left(  z\right)   &  =\frac{Z_{n}^{\pm}}{\sqrt{2}}\left(
z\frac{d}{dz}\mp z+1+n\right)  \frac{I_{n+1/2}\left(  z\right)  }{\sqrt{z}}\\
\psi_{n}^{\pm}\left(  z\right)   &  =Z_{n}^{\pm}\left(  \frac{z}{2}\right)
^{n}\left(  \sum_{k=0}^{\infty}\frac{1}{k!\Gamma\left(  k+\frac{1}%
{2}+n\right)  }\left(  \frac{z}{2}\right)  ^{2k}\mp\sum_{k=0}^{\infty}\frac
{1}{k!\Gamma\left(  k+\frac{3}{2}+n\right)  }\left(  \frac{z}{2}\right)
^{2k+1}\right)
\end{align}
If we choose $Z_{n}^{+}=Z_{n}^{-}$\ then the relations between the two sets of
eigenfunctions are just%
\begin{equation}
\left(  z\frac{d}{dz}\pm z\right)  \psi_{n}^{\pm}\left(  z\right)  =n\psi
_{n}^{\mp}\left(  z\right)  \label{MorseEigenfunctionEqn}%
\end{equation}
Now, what about the dual polynomials $\left\{  \chi_{n}^{\pm}\ |\ n\geq
0\right\}  $ which are the biorthonormalized duals for $\left\{  \psi_{n}%
^{\pm}\ |\ n\geq0\right\}  $? \ These are given by
\begin{equation}
\chi_{n}^{\pm}\left(  z\right)  =\frac{1}{Z_{n}^{\pm}}\left(  \frac{2}%
{z}\right)  ^{n}\left(  \sum_{k=0}^{\left\lfloor n/2\right\rfloor }%
\frac{\left(  -1\right)  ^{k}\Gamma\left(  n-k+\frac{1}{2}\right)  }%
{k!}\left(  \frac{z}{2}\right)  ^{2k}\pm\sum_{k=0}^{\left\lfloor \left(
n-1\right)  /2\right\rfloor }\frac{\left(  -1\right)  ^{k}\Gamma\left(
n-k-\frac{1}{2}\right)  }{k!}\left(  \frac{z}{2}\right)  ^{2k+1}\right)
\end{equation}
As illustration of the general theory, for all $k,n\geq0$ we have the
orthonormality and the completeness relation as given in (\ref{OrthoComplete}%
). \ It is straightforward to check these relations by using the explicit
series forms for $\chi_{n}^{\pm}$\ and $\psi_{n}^{\pm}$.

Again choosing both normalization factors to be the same, $Z_{n}^{+}=Z_{n}%
^{-}=Z_{n}$, the dual polynomials for the complex Morse potential are
solutions to the exceptionally simple inhomogeneous pair of equations%
\begin{equation}
\left(  z\frac{d}{dz}\pm z\right)  \chi_{n}^{\pm}\left(  z\right)  +n\chi
_{n}^{\mp}\left(  z\right)  =z\lambda_{n}^{\pm} \label{MorseDualPolyEqn}%
\end{equation}
where by direct calculation we find%
\begin{equation}
\lambda_{n}^{+}=\frac{\left(  -1\right)  ^{\left\lfloor n/2\right\rfloor }%
}{Z_{n}}\frac{\Gamma\left(  \left\lfloor n/2\right\rfloor +\frac{1}{2}\right)
}{\Gamma\left(  \left\lfloor n/2\right\rfloor +1\right)  }\ ,\ \ \ \lambda
_{n}^{-}=\left(  -1\right)  ^{n+1}\lambda_{n}^{+}%
\end{equation}
That is to say%
\begin{equation}
\lambda_{n}^{\pm}=\pm c_{n,n}^{\pm}%
\end{equation}
Moreover%
\begin{equation}
\left(  H_{\pm}-n^{2}\right)  \chi_{n}^{\pm}\left(  z\right)  =\left(
z\frac{d}{dz}\mp z\right)  z\lambda_{n}^{\pm}-zn\lambda_{n}^{\mp}=z\left(
\lambda_{n}^{\pm}-n\lambda_{n}^{\mp}\right)  \mp z^{2}\lambda_{n}^{\pm}%
\end{equation}
or with the explicit coefficients%
\begin{align}
\left(  H_{+}-n^{2}\right)  \chi_{n}^{+}\left(  z\right)   &  =\frac{\left(
-1\right)  ^{\left\lfloor n/2\right\rfloor }}{Z_{n}}\frac{\Gamma\left(
\left\lfloor n/2\right\rfloor +\frac{1}{2}\right)  }{\Gamma\left(
\left\lfloor n/2\right\rfloor +1\right)  }\left(  \left(  1+\left(  -1\right)
^{n}n\right)  z-z^{2}\right) \\
\left(  H_{-}-n^{2}\right)  \chi_{n}^{-}\left(  z\right)   &  =\frac{\left(
-1\right)  ^{n+1+\left\lfloor n/2\right\rfloor }}{Z_{n}}\frac{\Gamma\left(
\left\lfloor n/2\right\rfloor +\frac{1}{2}\right)  }{\Gamma\left(
\left\lfloor n/2\right\rfloor +1\right)  }\left(  \left(  1+\left(  -1\right)
^{n}n\right)  z+z^{2}\right)
\end{align}
The coefficients are a bit awkward, particularly the phases, but are dictated
by $\chi_{n}^{\pm}\left(  z\right)  \ _{\widetilde{_{z\rightarrow0}}}%
\ \frac{2^{n}\Gamma\left(  n+\frac{1}{2}\right)  }{Z_{n}}\frac{1}{z^{n}}$.

\paragraph{Singular potentials}

Now we go on to discuss models with several $\mu$s. \ In particular, if the
sums over $k$\ are infinite, the potentials can have fixed singularities for
finite values of $z$. \ We explore the situation for a particular
supersymmetric pair of such singular potentials. \ Namely those generated by
the superpotential%
\begin{equation}
U\left(  z\right)  =\frac{z}{1-z}%
\end{equation}
Up to the scale of $z$, this is the unique superpotential that reproduces
itself to obtain $V_{+}\left(  z\right)  =U\left(  z\right)  $.%
\begin{align}
V_{+}  &  =z\frac{d}{dz}\left(  \frac{z}{1-z}\right)  -\left(  \frac{z}%
{1-z}\right)  ^{2}=\frac{z}{1-z}=z+z^{2}+z^{3}+O\left(  z^{4}\right) \\
V_{-}  &  =-z\frac{d}{dz}\left(  \frac{z}{1-z}\right)  -\left(  \frac{z}%
{1-z}\right)  ^{2}=-z\frac{z+1}{\left(  1-z\right)  ^{2}}=-z-3z^{2}%
-5z^{3}+O\left(  z^{4}\right) \nonumber
\end{align}
What are the exact energy eigenfunctions for these potentials, analytic about
$z=0$? \ First consider the Hamiltonian with potential $V_{+}$.%
\begin{gather}
H_{+}=z^{2}\frac{d^{2}}{dz^{2}}+z\frac{d}{dz}+\frac{z}{1-z}=\left(  z\frac
{d}{dz}-\frac{z}{1-z}\right)  \left(  z\frac{d}{dz}+\frac{z}{1-z}\right)
\label{MorseFactorSingPlus}\\
z^{2}\frac{d^{2}}{dz^{2}}\psi_{n}^{+}+z\frac{d}{dz}\psi_{n}^{+}+\frac
{z}{\left(  1-z\right)  }\psi_{n}^{+}=n^{2}\psi_{n}^{+}\\
\psi_{n}^{+}\left(  z\right)  =z^{n}\left(  1-z\right)
\operatorname{hypergeom}\left(  \left[  1+n+\sqrt{1+n^{2}},1+n-\sqrt{1+n^{2}%
}\right]  ,\left[  1+2n\right]  ,z\right) \label{MorsePsi+}\\
\psi_{0}^{+}\left(  z\right)  =1-z\nonumber
\end{gather}
We note the especially simple form for the ground state. \ Excited states are
not such elementary functions. \ 

The dual polynomials in this case are%
\begin{align}
\chi_{n}^{+}\left(  z\right)   &  =\frac{1}{z^{n}}\left(  1+\sum_{k=1}%
^{n}\frac{z^{k}}{k!}\frac{\left(  2n-k-1\right)  !}{\left(  2n-1\right)
!}\frac{\Gamma\left(  k-n+\sqrt{n^{2}+1}\right)  \Gamma\left(  n+\sqrt
{n^{2}+1}\right)  }{\Gamma\left(  1-n+\sqrt{n^{2}+1}\right)  \Gamma\left(
-k+1+n+\sqrt{n^{2}+1}\right)  }\right) \label{MorseChi+}\\
\chi_{n}^{+}\left(  1\right)   &  =2\frac{\sqrt{1+n^{2}}}{\Gamma\left(
1+2n\right)  }\frac{\Gamma\left(  n+\sqrt{1+n^{2}}\right)  }{\Gamma\left(
1-n+\sqrt{1+n^{2}}\right)  } \label{MorseChi+1}%
\end{align}
These are solutions of the inhomogeneous equations%
\begin{equation}
\left(  H-n^{2}\right)  \chi_{n}^{+}\left(  z\right)  =V_{+}\left(  z\right)
\chi_{n}^{+}\left(  1\right)  =\frac{z}{1-z}\ \chi_{n}^{+}\left(  1\right)
\label{SingInhomo}%
\end{equation}
where the coefficient of the singular inhomogeneity is just $\chi_{n}^{+}$
evaluated at the singularity. \ The orthonormality relations between the
eigenfunctions and the dual polynomials are again the expected ones,
(\ref{OrthoComplete}). \ In this case, the contour encloses the origin once in
the positive counterclockwise sense, but lies within the unit-radius circle of
convergence of the series for $\psi_{n}^{+}\left(  z\right)  $. \ (Or at
least, the contour swerves \textquotedblleft to the left\textquotedblright\ to
avoid the singularity at $z=1$.) \ From the explicit series there also follows
the expression for the Cauchy kernel as given in (\ref{OrthoComplete}).

Now consider the Hamiltonian with the superpartner potential, $V_{-}$.%
\begin{gather}
H_{-}=z^{2}\frac{d^{2}}{dz^{2}}+z\frac{d}{dz}-z\frac{z+1}{\left(  1-z\right)
^{2}}=\left(  z\frac{d}{dz}+\frac{z}{1-z}\right)  \left(  z\frac{d}{dz}%
-\frac{z}{1-z}\right) \label{MorseFactorSingMinus}\\
z^{2}\frac{d^{2}}{dz^{2}}\psi_{n}^{-}+z\frac{d}{dz}\psi_{n}^{-}-z\frac
{z+1}{\left(  1-z\right)  ^{2}}\psi_{n}^{-}=n^{2}\psi_{n}^{-}\\
\psi_{n}^{-}\left(  z\right)  =z^{n}\left(  1-z\right)  ^{2}%
\operatorname{hypergeom}\left(  \left[  2+n+\sqrt{1+n^{2}},2+n-\sqrt{1+n^{2}%
}\right]  ,\left[  1+2n\right]  ,z\right) \label{MorsePsi-}\\
\psi_{0}^{-}\left(  z\right)  =\frac{1}{1-z}\nonumber
\end{gather}
Again, we note the especially simple form for the ground state. \ And again,
excited states are not such elementary functions. \ The dual polynomials are
now given by%
\begin{align}
\chi_{0}^{-}\left(  z\right)   &  =1\text{ \ \ and for \ \ }n>0\text{ \ \ }\\
\chi_{n}^{-}\left(  z\right)   &  =-\frac{1}{n}\left(  z\frac{d}{dz}\chi
_{n}^{+}\left(  z\right)  +\frac{z}{1-z}\left(  \chi_{n}^{+}\left(  z\right)
-\chi_{n}^{+}\left(  1\right)  \right)  \right)  \ \nonumber
\end{align}
After simplification of this last expression, we obtain%
\begin{align}
\chi_{n}^{-}\left(  z\right)   &  =\frac{1}{z^{n}}\left(  1-\frac{1}{n}%
\sum_{k=1}^{n}\frac{z^{k}}{k!}\left(  k-n+2nk-k^{2}\right)  \frac{\left(
2n-k-1\right)  !}{\left(  2n-1\right)  !}\frac{\Gamma\left(  k-n+\sqrt
{n^{2}+1}\right)  \Gamma\left(  n+\sqrt{n^{2}+1}\right)  }{\Gamma\left(
1-n+\sqrt{n^{2}+1}\right)  \Gamma\left(  -k+1+n+\sqrt{n^{2}+1}\right)
}\right) \nonumber\\
&  =\frac{\left(  1-z\right)  }{z^{n}}\sum_{j=0}^{n-1}z^{j}\frac{2\left(
n-j\right)  \Gamma\left(  2n-j\right)  }{j!\left(  2n\right)  !}\frac
{\Gamma\left(  1+j-n+\sqrt{n^{2}+1}\right)  \Gamma\left(  n+\sqrt{n^{2}%
+1}\right)  }{\Gamma\left(  1-n+\sqrt{n^{2}+1}\right)  \Gamma\left(
-j+n+\sqrt{n^{2}+1}\right)  } \label{MorseChi-}%
\end{align}
Note that $\chi_{n}^{-}\left(  1\right)  =0$. \ Just as (\ref{MorsePsi-}) has
an additional factor of $\left(  1-z\right)  $ compared to (\ref{MorsePsi+}),
so too does (\ref{MorseChi-}) compared to (\ref{MorseChi+}).

These results also exhibit biorthonormality and lead to yet another expression
for the Cauchy kernel, as in (\ref{OrthoComplete}), as may be established from
the explicit series. \ The inhomogeneous equation obeyed by $\chi_{0}^{-}$ is
obviously just
\begin{equation}
H_{-}\chi_{0}^{-}=V_{-}\chi_{0}^{-}=-z\frac{z+1}{\left(  1-z\right)  ^{2}}%
\end{equation}
On the other hand, the inhomogeneous equations obeyed by $\chi_{n>0}^{-}$ are
a bit more interesting.%
\begin{equation}
\left(  H-n^{2}\right)  \chi_{n}^{-}\left(  z\right)  =V_{+}\left(  z\right)
2\chi_{n}^{-\prime}\left(  1\right)  =\frac{z}{1-z}\left[  2\frac{d}{dz}%
\chi_{n}^{-}\right\vert _{z=1}%
\end{equation}
This is the same singular inhomogeneity as appears in the superpartner dual
polynomial equation, (\ref{SingInhomo}), only with a different coefficient.
\ In fact,%
\begin{equation}
2\frac{d}{dz}\chi_{n}^{-}\left(  1\right)  =-n\chi_{n}^{+}\left(  1\right)
\label{MorseChi-'1}%
\end{equation}
All this is more transparent upon exploiting the Darboux factorizations of the
Hamiltonians, (\ref{MorseFactorSingPlus}) and (\ref{MorseFactorSingMinus}).
\ As expected from the general theory, we have%
\begin{align}
\left(  H_{\pm}-n^{2}\right)  \psi_{n}^{\pm}\left(  z\right)   &  =0\\
\left(  H_{+}-n^{2}\right)  \chi_{n}^{+}\left(  z\right)   &  =V_{0}\left(
z\right)  \chi_{n}^{+}\left(  1\right)  =\frac{z}{1-z}\chi_{n}^{+}\left(
1\right) \\
\left(  H_{-}-n^{2}\right)  \chi_{n}^{-}\left(  z\right)   &  =-V_{0}\left(
z\right)  n\chi_{n}^{+}\left(  1\right)  =\frac{z}{1-z}2\chi_{n}^{-\prime
}\left(  1\right)
\end{align}
bearing in mind (\ref{MorseChi+1}) and (\ref{MorseChi-'1}), as well as%
\begin{align}
\left(  z\frac{d}{dz}\pm\frac{z}{1-z}\right)  \psi_{n}^{\pm}\left(  z\right)
&  =n\psi_{n}^{\mp}\left(  z\right) \\
\left(  -z\frac{d}{dz}-\frac{z}{1-z}\right)  \chi_{n}^{+}\left(  z\right)   &
=n\chi_{n}^{-}\left(  z\right)  -\frac{z}{1-z}\ \chi_{n}^{+}\left(  1\right)
\\
\left(  -z\frac{d}{dz}+\frac{z}{1-z}\right)  \chi_{n}^{-}\left(  z\right)   &
=n\chi_{n}^{+}\left(  z\right)  +\frac{z}{1-z}\delta_{n0}%
\end{align}
The last of these is strikingly simpler than expected, exhibiting an
inhomogeneity only for the case $n=0$.

Then there are the complementary, but subsidiary, relations:%
\begin{equation}
\left(  -z\frac{d}{dz}\pm\frac{z}{1-z}\right)  \chi_{n}^{\pm}\left(  z\right)
=n\chi_{n}^{\mp}\left(  z\right)  -\eta_{n}^{\pm}\left(  z\right)
\end{equation}
which allow the functions $\eta_{n}^{\pm}\left(  z\right)  $ to be constructed
from (\ref{MorseChi-}) and (\ref{MorseChi+}). \ Differentiating again gives%
\begin{equation}
\left(  H_{\mp}-n^{2}\right)  \chi_{n}^{\pm}\left(  z\right)  =-n\eta_{n}%
^{\mp}\left(  z\right)  -\left(  -z\frac{d}{dz}\mp\frac{z}{1-z}\right)
\eta_{n}^{\pm}\left(  z\right)
\end{equation}

\section{Magnetic field effects}

Now consider $\nu\neq0$. \ For a given $U\left(  z\right)  \equiv\sum
_{k>0}\upsilon_{k}z^{k}$\ we define%
\begin{align}
H_{\pm}  &  =\left(  z\frac{d}{dz}+\nu\right)  ^{2}-U\left(  z\right)  ^{2}\pm
z\frac{d}{dz}U\left(  z\right)  =\left(  z\frac{d}{dz}+\nu\mp U\right)
\left(  z\frac{d}{dz}+\nu\pm U\right) \label{MagneticH}\\
\widetilde{H}_{\pm}  &  =\left(  z\frac{d}{dz}-\nu\right)  ^{2}-U\left(
z\right)  ^{2}\pm z\frac{d}{dz}U\left(  z\right)  =\left(  z\frac{d}{dz}%
-\nu\mp U\right)  \left(  z\frac{d}{dz}-\nu\pm U\right)
\label{MagneticHTilde}%
\end{align}
Obviously, $\widetilde{H}_{\pm}\left[  \nu\right]  =H_{\pm}\left[
-\nu\right]  $. \ We seek eigenfunctions and associated functions that have
particular analytic behavior near $z=0$. \ 

\paragraph{Energy eigenfunctions}

We look for pairs of functions $\psi_{n}^{\pm}\left(  z\right)  $ that satisfy
the equations
\begin{equation}
\left(  z\frac{d}{dz}+\nu\pm U\left(  z\right)  \right)  \psi_{n}^{\pm}\left(
z\right)  =\left(  n+\nu\right)  \psi_{n}^{\mp}\left(  z\right)
\end{equation}
with $\psi_{n}^{\pm}\left(  z\right)  \ _{\widetilde{z\rightarrow0}}\ z^{n}$.
\ Usually these are infinite series, but again, for a given $U\left(
z\right)  $, both functions $\psi_{n}^{\pm}\left(  z\right)  $\ are now
completely specified. \ We note that negative integer $n$ are now admissible,
and independent of the corresponding $\left\vert n\right\vert $, for generic
$\nu$. \ It follows that the $\psi_{n}^{\pm}\left(  z\right)  $ are
eigenfunctions of $H_{\pm}$ with eigenvalues $\left(  n+\nu\right)  ^{2}$,
\begin{equation}
H_{\pm}\psi_{n}^{\pm}\left(  z\right)  =\left(  n+\nu\right)  ^{2}\psi
_{n}^{\pm}\left(  z\right)
\end{equation}
It also follows for generic $\nu$\ that a suitable dual to $\psi_{n}^{\pm
}\left(  z\right)  $ is provided by $\widetilde{\psi}_{-n}^{\pm}\left(
z\right)  $, where $\widetilde{H}_{\pm}\widetilde{\psi}_{-n}^{\pm}=\left(
-n-\nu\right)  ^{2}\widetilde{\psi}_{-n}^{\pm}$ so that $\widetilde{\psi}%
_{-n}^{\pm}$\ has the same energy as $\psi_{n}^{\pm}$. \ Nevertheless, as we
discuss shortly (cf. remarks about the \textquotedblleft
right-sector\textquotedblright\ given below), if we consider subsectors of the
spectrum, we may use polynomials as alternative dual functions, just as in the
periodic situations discussed above for $\nu\neq0$.

For example, reconsider the singular potential $U=\frac{z}{1-z}$. \ For
generic $\nu$ the solutions of%
\[
z^{2}\frac{d^{2}}{dz^{2}}f+\left(  1+2\nu\right)  z\frac{d}{dz}f+\nu
^{2}f+\frac{z}{1-z}f=\left(  n+\nu\right)  ^{2}f
\]
which are single-valued about $z=0$ are%
\begin{equation}
\psi_{n}^{+}\left(  z\right)  =z^{n}\left(  1-z\right)
\operatorname{hypergeom}\left(  \left[  1+n+\nu+\sqrt{1+\left(  n+\nu\right)
^{2}},1+n+\nu-\sqrt{1+\left(  n+\nu\right)  ^{2}}\right]  ,\left[
1+2n+2\nu\right]  ,z\right)
\end{equation}
where the indices on the hypergeometric function involve $\rho$, the roots of
$0=-1-n^{2}-2n\nu-2\nu\rho+\rho^{2}$. \ That is to say $\rho=\nu\pm
\sqrt{1+\left(  n+\nu\right)  ^{2}}$. \ The other solutions of the equation
are%
\[
z^{-n-2\nu}\left(  1-z\right)  \operatorname{hypergeom}\left(  \left[
-\rho+1-n,1+\rho-n-2\nu\right]  ,\left[  1-2n-2\nu\right]  ,z\right)
\]
but these have a branch point at $z=0$ and are not single-valued unless
$2\nu\in\mathbb{Z}$. \ 

For the partner potential, the single-valued solutions of%

\[
z^{2}\frac{d^{2}}{dz^{2}}f+\left(  1+2\nu\right)  z\frac{d}{dz}f+\nu
^{2}f-z\frac{z+1}{\left(  1-z\right)  ^{2}}f=\left(  n+\nu\right)  ^{2}f
\]
are%
\begin{equation}
\psi_{n}^{-}\left(  z\right)  =z^{n}\left(  1-z\right)  ^{2}%
\operatorname{hypergeom}\left(  \left[  2+n+\nu+\sqrt{1+\left(  n+\nu\right)
^{2}},2+n+\nu-\sqrt{1+\left(  n+\nu\right)  ^{2}}\right]  ,\left[
1+2n+2\nu\right]  ,z\right)
\end{equation}
while the solutions with a cut, for generic $\nu$, are%
\[
z^{-n-2\nu}\left(  1-z\right)  ^{2}\operatorname{hypergeom}\left(  \left[
2+\rho-n-2\nu,-\rho+2-n\right]  ,\left[  1-2n-2\nu\right]  ,z\right)
\]
where $\rho=\nu\pm\sqrt{1+\left(  n+\nu\right)  ^{2}}$ as before.

\paragraph{The dual polynomials and their inhomogeneities.}

On the \textquotedblleft right sector\textquotedblright\ $\left\{  \psi
_{n}^{\pm}\left(  z\right)  \ |\ n\geq0\right\}  $ (this terminology is
explained in \cite{CM}) we also look for pairs of \emph{finite dual
polynomials}\ in $z^{-1}$, of the form given in (\ref{DualPolys}). \ These
satisfy the inhomogeneous equations
\begin{equation}
\left(  z\frac{d}{dz}-\nu\pm U\left(  z\right)  \right)  \chi_{n}^{\pm}\left(
z\right)  +\left(  n+\nu\right)  \chi_{n}^{\mp}\left(  z\right)  =\Lambda
_{n}^{\pm}\left(  z\right)  \equiv\sum_{k>0}\lambda_{n}^{\pm}\left(  k\right)
z^{k} \label{MagneticFirstOrder}%
\end{equation}
where $U\left(  z\right)  $ and $\Lambda_{n}^{\pm}\left(  z\right)  $ are all
analytic about the origin, where they vanish. \ A priori the $\Lambda_{n}%
^{\pm}\left(  z\right)  $ are not given, but are to be determined along with
$\chi_{n}^{\pm}\left(  z\right)  $. \ Once more we normalize these polynomials
so that $\chi_{n}^{\pm}\left(  z\right)  \ _{\widetilde{z\rightarrow0}%
}\ 1/z^{n}$. \ This choice also fixes the normalization of $\Lambda_{n}^{\pm}%
$. \ For a given $U\left(  z\right)  $ the four functions $\chi_{n}^{\pm
}\left(  z\right)  $ and $\Lambda_{n}^{\pm}\left(  z\right)  $\ are now
completely specified.

The associated polynomials also obey the inhomogeneous Hamiltonian equations%
\begin{gather}
\widetilde{H}_{\pm}\chi_{n}^{\pm}\left(  z\right)  =\left(  z\frac{d}{dz}%
-\nu\mp U\right)  \left(  z\frac{d}{dz}-\nu\pm U\right)  \chi_{n}^{\pm}\left(
z\right)  =\left(  z\frac{d}{dz}-\nu\mp U\right)  \left(  \Lambda_{n}^{\pm
}\left(  z\right)  -\left(  n+\nu\right)  \chi_{n}^{\mp}\left(  z\right)
\right) \\
=\left(  z\frac{d}{dz}-\nu\mp U\right)  \Lambda_{n}^{\pm}\left(  z\right)
-\left(  n+\nu\right)  \left(  \Lambda_{n}^{\mp}\left(  z\right)  -\left(
n+\nu\right)  \chi_{n}^{\pm}\left(  z\right)  \right)
\end{gather}
That is to say%
\begin{equation}
\left(  \widetilde{H}_{\pm}-\left(  n+\nu\right)  ^{2}\right)  \chi_{n}^{\pm
}\left(  z\right)  =\left(  z\frac{d}{dz}-\nu\mp U\left(  z\right)  \right)
\Lambda_{n}^{\pm}\left(  z\right)  -\left(  n+\nu\right)  \Lambda_{n}^{\mp
}\left(  z\right)
\end{equation}
Note the RHS of this last equation involves only positive powers, $z^{n}$ for
$n\geq1$. \ 

Moreover, it also follows that for $k,n\geq0$\ and $-1/2<\nu<1/2$ these
functions are biorthonormal systems as in (\ref{OrthoComplete}). \ These
relations once again encode orthonormality and completeness on functions
analytic around $z=0$. \ For example, it is straightforward to show $\left(
n+\nu\right)  ^{2}\oint\frac{dz}{z}\ \chi_{k}^{\pm}\left(  z\right)  \psi
_{n}^{\pm}\left(  z\right)  =\left(  k+\nu\right)  ^{2}\oint\frac{dz}{z}%
\ \psi_{n}^{\pm}\left(  z\right)  \chi_{k}^{\pm}\left(  z\right)  $ by
inserting the Hamiltonian (\ref{MagneticH}) and integrating by parts to
convert to (\ref{MagneticHTilde}). \ That is%
\begin{equation}
0=\left(  \left(  k+\nu\right)  ^{2}-\left(  n+\nu\right)  ^{2}\right)
\oint\frac{dz}{z}\ \chi_{k}^{\pm}\left(  z\right)  \psi_{n}^{\pm}\left(
z\right)  =\left(  k-n\right)  \left(  k+n+2\nu\right)  \oint\frac{dz}%
{z}\ \chi_{k}^{\pm}\left(  z\right)  \psi_{n}^{\pm}\left(  z\right)
\end{equation}
For $k\neq n$ and both $\geq0$, and $2\nu\notin\mathbb{Z}_{<0}$, we conclude
$\oint\frac{dz}{z}\ \chi_{k}^{\pm}\left(  z\right)  \psi_{n}^{\pm}\left(
z\right)  =0$.

Once again the various functions are uniquely determined, given that\ $\chi
_{n}^{\pm}\left(  z\right)  \ _{\widetilde{z\rightarrow0}}\ 1/z^{n}$. \ We
choose the coefficients in $\chi_{n}^{\pm}\left(  z\right)  $ to eliminate all
$z^{-k}$ terms in (\ref{MagneticFirstOrder}), for $k=0,1,\cdots,n$, and then
we sweep all the remaining positive powers of $z$ into the $\Lambda_{n}^{\pm
}\left(  z\right)  $. \
\begin{gather}
\left(  z\frac{d}{dz}-\nu\pm U\left(  z\right)  \right)  \chi_{n}^{\pm}\left(
z\right)  +\left(  n+\nu\right)  \chi_{n}^{\mp}\left(  z\right)  =\Lambda
_{n}^{\pm}\left(  z\right) \\
U\left(  z\right)  \equiv\sum_{k>0}\upsilon_{k}z^{k}\ ,\ \ \ \chi_{n}^{\pm
}\left(  z\right)  \equiv\frac{1}{z^{n}}\sum_{j=0}^{n}c_{n,j}^{\pm}%
z^{j}\ ,\ \ \ \Lambda_{n}^{\pm}\left(  z\right)  \equiv\sum_{k>0}\lambda
_{n}^{\pm}\left(  k\right)  z^{k}\\
\sum_{j=0}^{n}c_{n,j}^{\pm}\left(  j-n-\nu\right)  z^{j}\pm\sum_{k>0}%
\upsilon_{k}z^{k}\sum_{j=0}^{n}c_{n,j}^{\pm}z^{j}+\left(  n+\nu\right)
\sum_{j=0}^{n}c_{n,j}^{\mp}z^{j}=\sum_{k>0}\lambda_{n}^{\pm}\left(  k\right)
z^{k+n}%
\end{gather}
So then%
\begin{gather}
c_{n,j}^{\pm}\left(  j-n-\nu\right)  \pm\sum_{k=1}^{j}\upsilon_{k}%
c_{n,j-k}^{\pm}+\left(  n+\nu\right)  c_{n,j}^{\mp}=0\text{ \ \ for
\ \ }j=0,\cdots,n\\
\text{and \ \ }\lambda_{n}^{\pm}\left(  k\right)  =\pm\sum_{j=0}^{n}%
\upsilon_{k+n-j}c_{n,j}^{\pm} \label{SuperLambdas}%
\end{gather}
Therefore the recursion goes like this. \ First we determine all the
coefficients in the $\chi_{n}^{\pm}\left(  z\right)  $ from the pair of
equations%
\[
\left(
\begin{array}
[c]{cc}%
\nu+n-j & -n-\nu\\
n+\nu & j-n-\nu
\end{array}
\right)  \left(
\begin{array}
[c]{c}%
c_{n,j}^{+}\\
c_{n,j}^{-}%
\end{array}
\right)  =\sum_{k=1}^{j}\upsilon_{k}\left(
\begin{array}
[c]{c}%
c_{n,j-k}^{+}\\
c_{n,j-k}^{-}%
\end{array}
\right)  \text{ \ \ for \ \ }j=0,\cdots,n
\]
where $c_{n,0}^{\pm}\equiv1$. \ Now $\det\left(
\begin{array}
[c]{cc}%
\nu+n-j & -n-\nu\\
n+\nu & j-n-\nu
\end{array}
\right)  =j\left(  2n+2\nu-j\right)  $ and $\left(
\begin{array}
[c]{cc}%
\nu+n-j & -n-\nu\\
n+\nu & j-n-\nu
\end{array}
\right)  ^{-1}=\frac{1}{j\left(  2n+2\nu-j\right)  }\left(
\begin{array}
[c]{cc}%
j-n-\nu & n+\nu\\
-n-\nu & \nu+n-j
\end{array}
\right)  $, so we have the recursion relations%
\begin{equation}
c_{n,j}^{\pm}=\frac{\pm1}{j\left(  2n+2\nu-j\right)  }\sum_{k=1}^{j}%
\upsilon_{k}\left(  \left(  j-n-\nu\right)  c_{n,j-k}^{\pm}+\left(
n+\nu\right)  c_{n,j-k}^{\mp}\right)
\end{equation}
Each $c_{n,j}^{\pm}$ depends on only the first $j$ coefficients in the
expansion for $U\left(  z\right)  $, i.e. just on $\upsilon_{k\leq j}$. \ Then
from the $c_{n,j}^{\pm}$\ we determine the $\lambda_{n}^{\pm}\left(  k\right)
$\ using (\ref{SuperLambdas}). \ Note that $\lambda_{n}^{\pm}\left(  k\right)
\neq0$ for $k>n$ is \emph{possible} here, depending on the values of the
$\upsilon$s. \ For example, for $n=2$:%
\begin{align}
c_{2,0}^{\pm}  &  =1\ ,\ \ \ c_{2,1}^{\pm}=\frac{\pm1}{2\nu+3}\upsilon
_{1}\ ,\ \ \ c_{2,2}^{\pm}=\frac{\pm1}{2\left(  \nu+1\right)  }\upsilon
_{2}-\frac{1}{2\left(  2\nu+3\right)  }\upsilon_{1}^{2}\\
\pm\lambda_{2}^{\pm}\left(  k\right)   &  =\upsilon_{k+2}+\upsilon
_{k+1}c_{2,1}^{\pm}+\upsilon_{k}c_{2,2}^{\pm}\nonumber\\
&  =\upsilon_{k+2}\pm\frac{1}{2\nu+3}\upsilon_{1}\upsilon_{k+1}+\left(
\frac{\pm1}{2\left(  \nu+1\right)  }\upsilon_{2}-\frac{1}{2\left(
2\nu+3\right)  }\upsilon_{1}^{2}\right)  \upsilon_{k}%
\end{align}
And for $n=3$: \ %

\begin{align}
c_{3,0}^{\pm}  &  =1\ ,\ \ \ c_{3,1}^{\pm}=\frac{\pm1}{2\nu+5}\upsilon
_{1}\ ,\ \ \ c_{3,2}^{\pm}=\frac{\pm1}{2\left(  \nu+2\right)  }\upsilon
_{2}-\frac{1}{2\left(  2\nu+5\right)  }\upsilon_{1}^{2}\ ,\ \ \ \\
c_{3,3}^{\pm}  &  =\frac{\pm1}{2\nu+3}\upsilon_{3}-\frac{\left(
4\nu+9\right)  }{6\left(  \nu+2\right)  \left(  2\nu+5\right)  }\upsilon
_{2}\upsilon_{1}\mp\frac{1}{2\left(  2\nu+3\right)  \left(  2\nu+5\right)
}\upsilon_{1}^{3}\\
\pm\lambda_{3}^{\pm}\left(  k\right)   &  =\upsilon_{k+3}+\upsilon
_{k+2}c_{3,1}^{\pm}+\upsilon_{k+1}c_{3,2}^{\pm}+\upsilon_{k}c_{3,3}^{\pm
}\nonumber\\
&  =\upsilon_{k+3}\pm\frac{1}{2\nu+5}\upsilon_{1}\upsilon_{k+2}+\left(
\frac{\pm1}{2\left(  \nu+2\right)  }\upsilon_{2}-\frac{1}{2\left(
2\nu+5\right)  }\upsilon_{1}^{2}\right)  \upsilon_{k+1}\nonumber\\
&  +\left(  \frac{\pm1}{2\nu+3}\upsilon_{3}-\frac{\left(  4\nu+9\right)
}{6\left(  \nu+2\right)  \left(  2\nu+5\right)  }\upsilon_{2}\upsilon_{1}%
\mp\frac{1}{2\left(  2\nu+3\right)  \left(  2\nu+5\right)  }\upsilon_{1}%
^{3}\right)  \upsilon_{k}%
\end{align}
Thus we have the lowest four dual polynomials and their inhomogeneities.%
\begin{align}
\chi_{0}^{\pm}\left(  z\right)   &  =1\ ,\ \ \ \Lambda_{0}^{\pm}\left(
z\right)  =\pm U\left(  z\right)  =\pm\sum_{k>0}\upsilon_{k}z^{k}\\
\chi_{1}^{\pm}\left(  z\right)   &  =\frac{1}{z}\pm\frac{1}{\nu+1}\upsilon
_{1}\ ,\ \ \ \Lambda_{1}^{\pm}\left(  z\right)  =\sum_{k>0}\left(  \frac
{1}{\nu+1}\upsilon_{1}\upsilon_{k}\pm\upsilon_{k+1}\right)  z^{k}\\
\chi_{2}^{\pm}\left(  z\right)   &  =\frac{1}{z^{2}}\pm\frac{1}{2\nu
+3}\upsilon_{1}\frac{1}{z}\pm\frac{1}{2\left(  \nu+1\right)  }\upsilon
_{2}-\frac{1}{2\left(  2\nu+3\right)  }\upsilon_{1}^{2}\\
\Lambda_{2}^{\pm}\left(  z\right)   &  =\sum_{k>0}\left(  \left(  \frac
{1}{2\left(  \nu+1\right)  }\upsilon_{2}\mp\frac{1}{2\left(  2\nu+3\right)
}\upsilon_{1}^{2}\right)  \upsilon_{k}+\frac{1}{2\nu+3}\upsilon_{1}%
\upsilon_{k+1}\pm\upsilon_{k+2}\right)  z^{k}%
\end{align}%
\begin{align}
\chi_{3}^{\pm}\left(  z\right)   &  =\frac{1}{z^{3}}\pm\frac{1}{2\nu
+5}\upsilon_{1}\frac{1}{z^{2}}+\left(  \pm\frac{1}{2\left(  \nu+2\right)
}\upsilon_{2}-\frac{1}{2\left(  2\nu+5\right)  }\upsilon_{1}^{2}\right)
\frac{1}{z}\nonumber\\
&  \pm\frac{1}{2\nu+3}\upsilon_{3}-\frac{\left(  4\nu+9\right)  }{6\left(
\nu+2\right)  \left(  2\nu+5\right)  }\upsilon_{2}\upsilon_{1}\mp\frac
{1}{2\left(  2\nu+3\right)  \left(  2\nu+5\right)  }\upsilon_{1}^{3}\\
\Lambda_{3}^{\pm}\left(  z\right)   &  =\sum_{k>0}\left(
\begin{array}
[c]{c}%
\left(  \frac{1}{2\nu+3}\upsilon_{3}\mp\frac{\left(  4\nu+9\right)  }{6\left(
\nu+2\right)  \left(  2\nu+5\right)  }\upsilon_{2}\upsilon_{1}-\frac
{1}{2\left(  2\nu+3\right)  \left(  2\nu+5\right)  }\upsilon_{1}^{3}\right)
\upsilon_{k}\\
+\left(  \frac{1}{2\left(  \nu+2\right)  }\upsilon_{2}\mp\frac{1}{2\left(
2\nu+5\right)  }\upsilon_{1}^{2}\right)  \upsilon_{k+1}+\frac{1}{2\nu
+5}\upsilon_{1}\upsilon_{k+2}\pm\upsilon_{k+3}%
\end{array}
\right)  z^{k}%
\end{align}
etc. \ Note that $\chi_{n}^{-}\left(  z\right)  $ and $\Lambda_{n}^{-}\left(
z\right)  $ are obtained from $\chi_{n}^{+}\left(  z\right)  $ and
$\Lambda_{n}^{+}\left(  z\right)  $, or vice versa, just by flipping the signs
of all the $\upsilon$s. \ These results reduce to those in the previous Table,
upon setting $\nu=0$. \ \ \ 

\paragraph{The partition function is not analytic in $\nu$}

For generic $\nu$%
\begin{align}
\mathcal{Z}\left[  \nu\right]   &  =\mathrm{trace}\left(  e^{-H\left[
\nu\right]  }\right)  =\sum_{n=-\infty}^{\infty}e^{-\left(  n+\nu\right)
^{2}}\nonumber\\
&  =e^{-\nu^{2}}\left(  \vartheta\left[  \nu\right]  +\vartheta\left[
-\nu\right]  -1\right)
\end{align}
upon defining the theta function\footnote{Or in terms of Jacobi's functions,
$\vartheta_{3}\left(  z,q\right)  \equiv\sum_{n=-\infty}^{\infty}q^{n^{2}%
}e^{2inz}$, so $\vartheta\left[  \nu\right]  +\vartheta\left[  -\nu\right]
-1=\vartheta_{3}\left(  z=i\nu,q=1/e\right)  $.}%
\begin{equation}
\vartheta\left[  \nu\right]  \equiv\sum_{n=0}^{\infty}e^{-n^{2}-2n\nu}%
\end{equation}
whereas for $\nu=0$, the partition function is%
\begin{equation}
\mathcal{Z}_{0}=\sum_{n=0}^{\infty}e^{-n^{2}}=\vartheta\left[  0\right]
=1.3863\cdots.
\end{equation}
So then%
\begin{equation}
\lim_{\nu\rightarrow0}\mathcal{Z}\left[  \nu\right]  =2\mathcal{Z}_{0}%
-1\neq\mathcal{Z}_{0}%
\end{equation}
This would suggest a phase transition as an indicator for bulk systems
governed by these dynamics.

\section{Conclusions}

We discussed supersymmetric biorthogonal quantum systems along the lines of
\cite{CM}, paying particular attention to the structure of non-hermitian
systems with periodic solutions, for which cases the duals of the energy
eigenfunctions are not simply related to the eigenfunctions by either complex
conjugation or PT reflection. \ We worked out the general theory for single
particle quantum systems, and we illustrated the general theory with several
explicit exact examples. \ 

It remains to investigate many-body or field theoretic extensions of these
supersymmetric systems, say by adapting the perturbative methods in \cite{CG}
on supersymmetric Liouville field theory, or by employing the powerful
non-perturbative methods of conformal field theory \cite{CFTbook}. \ This
additional study is in progress, and represents one application of the
formalism presented in this paper. \ There is a rich literature on Liouville
and super-Liouville theory, models whose importance came to light in the work
of Polyakov on string theory \cite{PolyakovString}, but which were
subsequently developed much further in the context of conformal field theory
and its applications to critical phenomena, as well as to subcritical string
theory \cite{LiouvilleReview}. \ In particular, the super-Liouville
correlation functions \cite{SuperLiouvilleCorrelators}\ have been shown to
exhibit interesting analytic behavior in the exponential coupling constant,
similar to the analytic structure of correlators for non-supersymmetric
Liouville field theory \cite{LiouvilleCorrelators}. \ The behavior of these
correlators is related to properties of various WZNW models
\cite{LiouvilleWZNW}, and there are particularly intriguing features that
correspond to purely imaginary coupling constants -- precisely the field
theory extensions of the type of models discussed in this paper.

As for other applications of supersymmetric biorthogonal quantum systems, say
to non-relativistic situations, an interesting possibility would be to
consider driven/dissipative condensates as suggested in \cite{CM}, but with
additional fermions in the condensate \cite{NRSuperStringCondensate}. \ This
too is under study.

\paragraph{Acknowledgements}

We thank C Bender for introducing us to PT symmetric theories and raising our
interest in problems involving non-hermitian Hamiltonians. \ We also thank P G
O Freund and A Veitia for useful discussions. \ One of us (TC) thanks the
Aspen Center for Physics for providing the stimulating environment in which
parts of this investigation were carried out during June - July 2005, and he
also thanks the Institute for Advanced Study, where this work was completed,
for its hospitality and support as a visiting Member during January - July
2006. \ This material is based upon work supported by the National Science
Foundation under Grant No's. 0303550 and 0555603.

\end{document}